\documentclass[showpacs,preprintnumbers,amsmath,amssymb]{revtex4}
\usepackage{bm}

\newcommand{\be}{\begin{equation}}
\newcommand{\ee}{\end{equation}}
\newcommand{\bea}{\begin{eqnarray}}
\newcommand{\eea}{\end{eqnarray}}
\newcommand{\nn}{\nonumber}
\newcommand{\p}{\phi_0}
\newcommand{\rd}{\partial}
\newcommand{\pp}{\tilde \phi}
\newcommand {\pb}{\bar \phi}

\begin{document}
\title{Fubini vacua as a classical de Sitter vacua}
\author{F. Loran}
\email{loran@cc.iut.ac.ir}

 \affiliation{Department of  Physics, Isfahan University of Technology
 (IUT), Isfahan,  Iran}

 \begin{abstract} The Fubini's idea to introduce
 a fundamental scale of hadron phenomena by means of dilatation
 non-invariant vacuum state in the frame work of a scale invariant
 Lagrangian field theory is recalled. The Fubini vacua is invariant
 under the  de~Sitter subgroup of the full conformal group. We obtain a finite entropy for the quantum state corresponding to the classical
 Fubini vacua in Euclidean space-time resembeling the entropy of the de~Sitter vacua. In Minkowski space-time it is shown that
 the Fubini vacua is mainly a bath of radiation with Rayleigh-Jeans
 distribution for the low energy radiation. In four dimensions, the critical scalar theory is
 shown to be equivalent to the Einstein field equation in the ansatz
 of conformally  flat metrics and to the $SU(2)$ Yang-Mills theory
 in the 't Hooft ansatz. In D-dimensions, the Hitchin formula for the information geometry metric of the moduli space of instantons is used
 to  obtain the information geometry of
 the free-parameter space of the Fubini vacua which is shown to be a $(D+1)$-dimensional AdS space. Considering the Fubini vacua as a de~Sitter vacua,
 the corresponding  cosmological constant is shown to be given by the coupling constant of the critical scalar theory. In Minkowski spacetime it is shown
 that the Fubini vacua is equivalent to an open FRW universe.
 \end{abstract}

 \pacs{98.80.-k,04.62.+v}
 \maketitle

\section{Introduction}
 The WMAP results \cite{WMAP} combined with earlier cosmological observations
 shows that we are living in an accelerating universe. The currently observed lumpiness in the temperature
 of the cosmic microwave background is just right for a flat
 universe though there are also some evidences that our universe
 is spatially open \cite{Gott}. The great simplifying fact of cosmology is that the universe appears to be homogeneous
 and isotropic along a preferred set of spatial hypersurfaces \cite{16}. Of course
 homogeneity and isotropy are only approximate, but they become
 increasingly good approximations on larger length scales, allowing
 us to describe spacetime on cosmological scales by the
 Robertson-Walker metric.

 Constructing four dimensional de Sitter vacuum from a field theory coupled to gravity
 has been a long standing challenge. Our main purpose in the present
 work is to give some evidences for considering the Fubini's approach
 as a possible solution to the problem.  In \cite{Fubini-paper},
 Fubini obtained a classical vacua of the critical scalar theory
 which preserves the de~Sitter subgroup of the full conformal
 symmetry of the theory. The idea was to introduce
 a fundamental scale of hadron phenomena by means of dilatation
 non-invariant vacuum state in the frame work of a scale invariant
 Lagrangian field theory. He verified that this program can only be carried out if the vacuum state is not
 translation invariant. In the following, among the other results, it is shown that the entropy of the Fubini vacua in Euclidean space resembles
 the conjectured entropy of the de~Sitter space and the cosmological
 constant of the de~Sitter vacua corresponding to the Fubini vacua is given classically by the coupling
 constant of the critical scalar theory. We think that these
 evidences are enough to consider and study the Fubini's
 method as a presumable solution for the problem of the
 cosmological constant.

 The present paper is organized as follows. A brief review of the Fubini's
 original work is given in section \ref{Fubini}, where we also discuss a possible generalization of
 the Fubini's approach applicable as dimensional reduction method, see section \ref{dimensional reduction}. The
 entropy of the Fubini vacua and its properties as a thermal bath is
 studied in section \ref{thermal}. The critical scalar theory and
 its relation to the Einstein field equation in the ansatz of
 conformally flat metrics and to $SU(2)$ Yang-Mills theory in the 't~Hooft ansatz
 is studied in section \ref{critical}. Specially in section \ref{'t
 Hooft} by applying the Hitchin formula, for a critical scalar theory in $D$-dimensions the information geometry
 metric of the free-parameter space of the Fubini classical vacua is shown to be a $(D+1)$-dimensional
 AdS space in which the Fubini classical solution appears as the
 boundary to bulk  propagator. Motivated by the Fubini's result we
 study the Fubini vacua as a de~Sitter vacua in section
 \ref{deSitter}. We show that the cosmological constant is
 proportional to the coupling constant of the critical theory. In
 section \ref{FRW} it is shown that in Minkowski spacetime, the Fubini
 vacua corresponds to an open FRW universe.
 Section \ref{conclusion} is devoted to conclusion and summarizing the results. In appendix \ref{Hitchin} the Hitchin formula for the information
 geometry metric of the moduli space of the Yang-Mills instantons
 which we apply to the Fubini classical vacua is reviewed. In
 appendix \ref{general aspect} the general properties of
 conformally flat geometries are discussed and the
 information about the Ricci and the scalar curvature tensors of
 such backgrounds necessary to read sections \ref{Einstein} and \ref{deSitter} are given.
 Finally in appendix \ref{stability} we study the Fubini classical
 vacua as a metastable stationary trajectory of the critical scalar
 theory.

 \section{Fubini vacua}\label{Fubini}
 In this section we give a brief review the Fubini approach to conformal
 invariant  field theories \cite{Fubini-paper}. The part of his work relevant to the present paper
 is the construction of a classical vacua for
 critical scalar theories that is invariant under the de~Sitter
 subgroup of the conformal group:

 ``{\em The physical idea is to introduce
 a fundamental scale of hadron phenomena by means of dilatation
 non-invariant vacuum state in the frame work of a scale invariant
 Lagrangian field theory. A new unconventional feature is that this
 program can only be carried out if the vacuum state is not
 translation invariant. The vacuum is still invariant under a
 10-parameter subgroup of the full conformal group ....}"

 We finally propose this vacua as a solution to the problem of cosmological constant. In section \ref{dimensional reduction} we discuss the
 usefulness of the Fubini approach (i.e. spontaneously breaking the translation symmetry) for dimensional reduction
 purposes.

 In $D$-dimensional Minkowski spacetime with Lorenz symmetry ${\cal O}(1,D-1)$ consider the conformal invariant Lagrangian
 \be
 {\cal L}=\frac{1}{2}(\rd_\mu\phi)(\rd^\mu\phi)-\frac{g}{\frac{2D}{D-2}}\phi^{\frac{2D}{D-2}}.
 \label{fu1}
 \ee
 One is looking for a solution of the field theoretical problem in
 which the vacuum expectation value of the field $\phi(x)$ is
 non-vanishing,
 \be
 \left<\Omega|\hat\phi|\Omega\right>=\p(x).
 \label{fu2}
 \ee
 $\p$ satisfies the equation of motion
 \be
 \Box \phi(x)+g\phi^{\frac{D+2}{D-2}}=0.
 \label{fu3}
 \ee
 In trying to solve Eq.(\ref{fu3}) one shall be guided by invariance
 considerations. The classical equation of motion (\ref{fu3}) is
 invariant under the full conformal group but it does not admit any
 non-trivial solution invariant under the full conformal group, as
 can be seen from the invariance condition,
 \be
 \left<\Omega|[{\cal G}_i,\phi(x)]|\Omega\right>=0,
 \label{fu4}
 \ee
 where ${\cal G}_i$'s are all conformal generators. If $\p$ is assumed to be translation invariant it should be a constant but for $g\neq
 0$, Eq.(\ref{fu3}) does not admit any constant solution.
 The Fubini's solution to Eq.(\ref{fu3}) is not invariant under
 translation but under the transformation generated by,
 \be
 R_\mu=\frac{1}{2}\left(\beta
 P_\mu+\frac{1}{\beta}K_\mu\right),
 \label{fu5}
 \ee
 in which $P_\mu$ is the translation generator and $K_\mu=I P_\mu
 I$, where $I$ is the operator of the inversion transformation
 $x_\mu\to x_\mu/x^2$. $\beta$ is some arbitrary constant and has
 dimension of a length. One can show that $R_\mu$ and the generators
 of the Lorenz group in $D$-dimensions together form an ${\cal O}(2,D-1)$
 or ${\cal O}(1,D)$ subgroup of the full conformal group ${\cal O}(2,D)$ if the
 coupling constant $g<0$ or $g>0$ respectively.\footnote{When comparing to the original work \cite{Fubini-paper} it should be noted that
 the Minkowski metric we use is $(-,+,\cdots,+)$ while Fubini assumes the metric $(+,-,\cdots,-)$.}
 Invariance of $\p$ under the generators of the Lorenz group implies
 that $\p=\p(x^2)$. Invariance under $R_\mu$ gives,
 \be
 \p=c\left(\frac{\beta^2+x^2}{2\beta}\right)^{-\left(\frac{D-2}{2}\right)},
 \label{fu6}
 \ee
 where $c$ is some constant to be determined by the equation of
 motion (\ref{fu3}). The final result is
 \be
 \p(\vec x)=\frac{\alpha}{\left(\beta^2+(\vec x-\vec a)^2\right)^{\frac{D-2}{2}}},\hspace{1cm}
 (\vec x-\vec a)^2=\delta_{\mu\nu} (x-a)^\mu(x-a)^\nu,
 \label{fu7}
 \ee
 where
 \be
 \alpha=\left(\frac{g}{D(D-2)\beta^2}\right)^{\frac{D-2}{4}},
 \label{fu8}
 \ee
 The free parameters $\beta$ and $a^\mu$ are present as far as the broken
 part of the conformal group contains $D+1$ generators corresponding
 to the dilatation operator which amounts to changing the choice of
 $\beta$ and translation operators which change $a^\mu$'s. In
 section \ref{'t Hooft} we will apply the Hitchin formula to obtain the
 information geometry of the free-parameter space. Interestingly one
 realizes that the information geometry is $\mbox{AdS}_{D+1}$ if
 $g>0$ and $\mbox{dS}_{D+1}$ if $g<0$.
 \subsection{Fubini's approach as a dimensional reduction method}\label{dimensional reduction}
 An interesting question is whether there exist a classical vacuum invariant under
 some translations if not all. In such a vacua all massless fields coupled to the scalar field become massive along those directions that
 translation symmetry is
 spontaneously broken while remain massless in other directions. In this way at low-energy limit one observes a lower dimensional space with Poincare
 symmetry as will be shown in this section.

 If we label directions along which
 the modified vacua $\p^m$ is invariant by $x_i$, $i=1,\cdots,D'$,
 and the remaining spatial directions (along which translation symmetry is spontaneously broken)
 by $x_a$, $a=1,\cdots,D-D'$,  then it is obvious that $\p^m$ is independent of
 $x_i$. From the conformal algebra,
 \bea
 [M_{\mu\nu},M_{\rho\sigma}]&=&-i\left(\eta_{\mu\rho}M_{\nu\sigma}-\eta_{\nu\rho}M_{\mu\sigma}+\eta_{\mu\sigma}M_{\rho\nu}-
 \eta_{\nu\sigma}M_{\rho\mu}\right),\nn\\
  \left[M_{\mu\nu},\left(\begin{array}{ll}P_\rho\\K_\rho\end{array}\right)\right]&=&
 i\left\{\eta_{\nu\rho}\left(\begin{array}{ll}P_\mu\\K_\mu\end{array}\right)-
 \eta_{\mu\rho}\left(\begin{array}{ll}P_\nu\\K_\nu\end{array}\right)\right\},\nn\\
 \left[{\cal D},\left(\begin{array}{ll}P_\rho\\K_\rho\end{array}\right)\right]&=&-i\left(\begin{array}{cc}P_\rho\\-K_\rho\end{array}\right),\nn\\
 \ [P_\mu,K_\nu]&=&2i(\eta_{\mu\nu}{\cal D}-M_{\mu\nu}),\nn\\
 \ [M_{\mu\nu},{\cal D}]&=&0, \hspace{1cm}[P_\mu,P_\nu]=[K_\mu,K_\nu]=0,
 \label{fu9}
 \eea
 and
 \bea
 \ [\phi(x),M_{\mu\nu}]&=&i(x_\mu\rd_\nu-x_\nu\rd_\mu)\phi(x),\nn\\
 \ [\phi(x),{\cal D}]&=&i\left(x^\mu\rd_\mu+\frac{D-2}{2}\right)\phi(x),\nn\\
 \ [\phi(x),P_\mu]&=&i\rd_\mu\phi(x),\nn\\
 \ [\phi(x),K_\mu]&=&i\left[-x^2\rd_\mu+2x_\mu\left(x^\rho\rd_\rho+\frac{D-2}{2}\right)\right]\phi(x),
 \label{fu10}
 \eea
 one easily verifies that a nontrivial solution $\phi(x_a)$ of equation of
 motion, can be assumed to be invariant under $M_{ij}$, ${\cal D}$ and
 $P_i$. The Lorentz generators $M_{ij}$ and the translation
 generators $P_i$ cover the Poincare symmetry group in $D'$
 dimension. Consequently a vacuum invariant under transformations
 generated by these generators breaks spontaneously the full conformal symmetry ${\cal
 O}(2,D)$ to Poincare group ${\cal O}(1,D'-1)$. The equation of motion for $\phi(x_a)$ is
 \be
 \Box_a \phi(x_a)+g\phi^{\frac{D+2}{D-2}}=0.
 \label{fu11}
 \ee
 To solve this equation one can assume invariance under $M_{ab}$
 which gives $\phi(x_a)=\phi(r_a)$ where $r_a^2=x_ax^a$.
 Invariance under dilatation $\cal D$, is not helpful, since
 \be
 [\phi(r_a),{\cal
 D}]=i\left(x^b\rd_b+\frac{D-2}{2}\right)\phi(r_a)=0,
 \label{fu12}
 \ee
 gives
 \be
 \phi(r_a)\sim r_a^{\frac{2-D}{2}}
 \label{fu13}
 \ee
 which is singular for $D>2$.
 The equation of motion for $\phi(x_a)=\phi(r_a)$ is
 \be
 \phi''+\frac{P-1}{r_a}\phi'+g\phi^{\frac{D+2}{D-2}}=0,
 \label{fu14}
 \ee
 where a $'$ denotes one time derivation with respect to $r_a$ and
 $P=D-D'$. We finish this section by giving a solution to this
 equation for $D=4$ and $P=1$. In this case the equation of motion
 is
 \be
 \phi''+g\phi^3=0,
 \label{fu15}
 \ee
 which can be easily integrated to give,
 \be
 \frac{1}{2}\phi'^2+\frac{g}{4}\phi^4=c.
 \label{fu16}
 \ee
 For $c=0$ the solution $\phi\sim r_a^{-1}$ is singular at $r_a=0$.
 For $c>0$, defining $c=L^{-4}$ one obtains,
 \be
 \phi=\frac{1}{L}\left(\frac{4}{g}\right)^{1/4}\mbox{sn}\left(\left.g^{1/4}\frac{r_a}{L}\right|-1\right),
 \label{fu17}
 \ee
 in which $\mbox{sn}(u|m)=\sin(\phi)$ is the Jacobi elliptic function  in which
 $\phi=\mbox{am}(u|m)$ is the inverse of Jacobi elliptic function of
 the first kind. For practical purposes one can assume
 $\mbox{sn}(x|-1)=\sin(1.2x)$ by a good precision.
 \section{Fubini vacua as a thermal bath}\label{thermal}
 In this section assuming a four dimensional spacetime we obtain the quantum state $\left|\Omega\right>$ corresponding to the
 classical vacua $\p$ and study it as thermal bath of free scalar  particles. In the Euclidean
 spacetime we show that the entropy $S$, given by the relation
 \be
 \left<\Omega|\Omega\right>=e^{-S},
 \label{tb1}
 \ee
 is a finite value,
 \be
 S\sim \frac{1}{g^{1/2}}\left(\frac{\ell}{\beta}\right)^2,
 \label{tb2}
 \ee
 in which $\ell$ is the radius of universe (the volume of box).
 In the Minkowski spacetime $\left|\Omega\right>$ is mainly a bath
 of radiation but a few percent of its content are tachyons and
 massive particles. Our method to construct $\left|\Omega\right>$ from the vacua $\left|0\right>$
 which is the empty-space quantum state
 annihilated by the annihilation operators of the whole (mass) spectrum of (scalar)
 particles, is to Fourier transform $\p$ to obtain the spectrum of
 plane waves superposed to construct it. We then create the same
 composition by operating the creation operators on $\left|0\right>$.
 \subsection{The entropy of Fubini vacua in Euclidean spacetime}
 Imagining the universe as a spherical box of radius $\ell$  we define a dimensionless parameter $x=r/\ell$. If the scalar fields are assumed to
 vanish on the boundary of the universe then $\ell$ should be large enough $\ell\gg \beta$ such that
 $\p$ be a good approximation of  the true classical vacua of the
 theory, since $\p$ vanishes at $r\to\infty$. Instead of $\p(r)$ which mass dimension
 in four dimensions is $[\p]=L^{-1}$ we work with the dimensionless
 field $\p(x)=\ell\p(r)$,
 \be
 \p(x)=\sqrt{\frac{8}{g}}\frac{\beta/\ell}{\left(\beta/\ell\right)^2+x^2}
 \label{tbe1}
 \ee
 By the Fourier transformation
 \be
 \phi(\vec x)=\int\tilde\phi(k)e^{i\vec k.\vec x}
 d^4k,
 \label{tbe2}
 \ee
 one can determine $\tilde\phi(k)$, i.e. the number of plane waves $e^{i\vec k.\vec
 x}$ with 4-momentum $\vec k$, constructing $\p$. By spherical
 symmetry of $\p$, an Euclidean observer living in the vacua $\p$ detects an
 isotropic radiation of free particles with spectrum
 $\tilde\phi(k)$. Up to some numerical constant $\tilde\phi(k)$ is
 given as follows,
 \be
 \tilde\phi(k)=\sqrt{\frac{8}{g}}\frac{\beta}{\ell}\frac{K_1(\frac{\beta}{\ell}\left|k\right|
 )}{\left|k\right|},
 \label{tbe3}
 \ee
 where $K_n(z)$ is the modified Bessel function of the second kind
 satisfying the differential equation
 \be
 z^2y''+z y'-(z^2+n^2)y=0.
 \label{tbe4}
 \ee
 Considering the annihilation and creation operators $a_k$ and $a^\dag_k$ satisfying the
 identities,
 \be
 \begin{array}{ccccc}
 a^\dag(k)
 \left|0\right>=\left|k\right>,&&a(k)\left|0\right>=0,&&[a(k),a^\dag(k')]=\delta^4(k-k'),
 \end{array}
 \label{tbe5}
 \ee
 in which $\left|0\right>$ is the vacua corresponding the empty
 spacetime with norm $\left<0|0\right>=1$,  we claim that the
 quantum state $\left|\Omega\right>$ corresponding to the classical
 vacua $\p$ is given by
 \be
 \left|\Omega\right>=\exp\left(\int
 d^4k{\tilde\phi(k)}^{1/2}a^\dag(k)\right)\left|0\right>,
 \label{tbe6}
 \ee
 since
 \be
 \frac{\left<\Omega|N(k)|\Omega\right>}{\left<\Omega|\Omega\right>}=\tilde\phi(k),
 \label{tbe7}
 \ee
 where $N(k)=a^\dag(k)a(k)$ is the number operator. The above result shows that number of plane waves $\left|k\right>$ that can be found in
 $\left|\Omega\right>$ is equal to the number of plane waves $e^{i\vec k.\vec x}$ in
 $\p$. It can be also shown that the state $\left|\Omega\right>$ also
 satisfies the Fubini's condition  $\left<\Omega|\hat\phi|\Omega\right>=\p$.\footnote{To show it one has to do some slight modifications. Consider
 $r$ as the time coordinate and define the Fourier transformation by $\p(r)=\int_{-\infty}^\infty dk e^{ikr}\tilde\phi(k)$.
 For second quantization consider $a(k)$ as the annihilation or creation operator for negative and positive values of $k$ respectively.
 The modification of definition (\ref{tbe6}) and  the algebra (\ref{tbe5}) is straightforward. Defining the operator
 $\hat\phi=\int_{-\infty}^\infty a(k) e^{ikr}$ it is easy to verify the Fubini condition (\ref{fu2}).}

 The entropy as given by the identity (\ref{tb1}) is given as
 follows,
 \be
 S\sim\int
 d^4k\tilde\phi(k)\sim\sqrt{\frac{8}{g}}\left(\frac{\ell}{\beta}\right)^2.
 \label{tbe8}
 \ee
 It is easy to show that in six dimensions the Fourier transform of
 $\phi(x)=\ell^2\p(r/\ell)$ is given by
 \be
 \tilde\phi_6(k)\sim\frac{1}{g}\left(\frac{\beta}{\ell}\right)^3\frac{K_1(\frac{\beta}{\ell}\left|k\right|)}{\left|k\right|},
 \label{tbe9}
 \ee
 and consequently the entropy of the corresponding quantum state is
 given by
 \be
 S_6\sim \frac{1}{g}\left(\frac{\ell}{\beta}\right)^4.
 \label{tbe10}
 \ee
 The Fourier transform of $\p$ in three dimensions can not be
 found analytically. But even by comparing the results given in
 Eqs.(\ref{tbe8}) and (\ref{tbe10}) by the entropy of the D-dimensional empty
 Euclidean de Sitter space
 \be
 S\sim \frac{\ell^{(D-2)}}{G_N},
 \label{tbe11}
 \ee
 a nice similarity can be realized.\footnote{In section {\ref{Einstein}} it is shown that  $\p$ gives a solution of the Einstein equation with
 positive cosmological constant. See also section \ref{deSitter}}
 \subsection{Spectrum of the Fubini  vacua in Minkowski spacetime}
 In Minkowski spacetime $\p$ has a singularity on the hypersurface
 given by $t^2-\vec x^2=\beta^2$. The Fourier transform of
 $\phi(\vec x,\tau)=\beta\p(\vec r,t)$ is given as follows,
 \bea
 \tilde\phi(\omega,\vec k)&\sim&\left(\frac{1}{g}\right)^{1/2}\beta^{-1}\int d^3x e^{-i \vec k. \vec x}\int
 d\tau\frac{e^{i\omega\tau}}{1-t^2+\vec x^2}\nn\\
 &=&\left(\frac{1}{g}\right)^{1/2}\beta^{-1}\int d^3x e^{-i \vec k. \vec x}\left[\frac{\pi}{\sqrt{1+\vec
 x^2}}\sin(\left|\omega\right|\sqrt{1+\vec x^2})\right]\nn\\
 &=&\left(\frac{1}{g}\right)^{1/2}\beta^{-1}\frac{(2\pi)^2}{k}\int_0^\infty dx\frac{x\sin(k
 x)\sin(\left|\omega\right|\sqrt{1+x^2})}{\sqrt{1+x^2}}.
 \label{tbm1}
 \eea
 The above integral is not convergent. The divergency can be
 analyzed as follows. If we split the integration interval
 $(0,\infty)$ into two subintervals $(0,l]$ and $(l,\infty)$ for
 some $l\gg 1$, one verifies that in the second interval the
 integrand is approximately equal to $\sin(k x)\sin(\left|\omega\right| x)$. Thus $\tilde\phi(\omega,\vec k)$ is approximately given by,
 \be
 k\tilde\phi(\omega,\vec k)\sim\int_0^l dx x\sin(k
 x)\left(\frac{\sin(\omega \sqrt{1+x^2})}{\sqrt{1+x^2}}-\frac{\sin
 (k x)}{x}\right)+\int_l^\ell dx \sin(k x)\sin(\omega x),
 \label{tbm2}
 \ee
 where $\ell\to\infty$ gives the volume of the box.
 $\tilde\phi(\omega,\vec k)$ is mainly given by the second term
 above which is approximately equal to $$\frac{1}{k}\left(\frac{\ell}{\pi/k}\right)\delta(\omega-k).$$

 In other words the classical vacuum $\p$ corresponds to radiation
 of massless scalars ($\omega=k$) and a few percent amount of various
 massive and tachyonic scalar fields ($\omega\neq k$). The details of the full spectrum is  given by
 Eq.(\ref{tbm1}). The distribution of massless scalar is given by the  formula,
 \be
 n(\omega) d\omega=d\omega\int d^3k\tilde\phi(\omega,\vec k)\sim
 \beta^{-1}\omega^2 d\omega,
 \label{tbm3}
 \ee
 which is the Rayleigh-Jeans formula for black-body radiation at
 temperature
 \be
 T=1/\beta.
 \label{tbm4}
 \ee
 The final result is in agreement with the black-body radiation.
 Because the Rayleigh-Jeans distribution formula gives the
 distribution of the black-body radiation for $\omega\ll T$. On the
 other hand we obtained Eq.(\ref{tbm3}) by an integration over
 $x>\beta l\gg\beta$ which corresponds to $k/\beta\ll\beta^{-1}$.
 \section{The critical scalar theory}\label{critical}
 In section \ref{Fubini} the Fubini vacua $\p$ as the classical vacua of the critical scalar
 theory, invariant under the de~Sitter subgroup of the conformal group  was
 obtained. In this section we at first give $\p$ as the solution of the
 nonlinear Laplace equation \cite{phi4},
 \be
 \sum_{i=1}^D\rd^2_i\phi+g\phi^{\frac{D+2}{D-2}}=0.
 \ee
 Then we show that in four
 dimensions the critical scalar theory corresponds to the Einstein
 gravity and $SU(2)$ Yang-Mills theory in special ansatz
 respectively and study $\p$ in those contexts.

 The Klein-Gordon equation for $SO(d+1)$-invariant solutions $\phi=\phi(r)$, where
 $r^2={x_\mu x^\mu}$ is
 \be
 \left(\frac{d^2}{dr^2}+\frac{d}{r}\frac{d}{d r}\right)\phi+g\phi^n=0.
 \label{d1}
 \ee
 One solution of this equation is
 \be
 \phi_s(r)=\left(\frac{2(d-1-\frac{2}{n-1})}{(n-1)}\right)^\frac{1}{n-1}
 \left(\frac{1}{g r^2}\right)^\frac{1}{n-1}
 \label{d2}
 \ee
 These solutions can be obtained by considering the ansatz  $\phi(r)=\alpha
 r^\beta$ and solving the wave equation to determine $\alpha$ and $\beta$.
 The above solutions become singular as $g\to 0$. One can show that
 Eq.(\ref{d1}) has also solutions like,
 \be
 \p(\vec x)=\frac{\alpha}{\left(\beta^2+(\vec x-\vec a)^2\right)^\gamma},\hspace{1cm}
 (\vec x-\vec a)^2=\delta_{\mu\nu} (x-a)^\mu(x-a)^\nu.
 \label{d3}
 \ee
 for some constants $a^\mu$, $\alpha$, $\beta$ and $\gamma$ only for conformally coupled theories, i.e. for $d=2,3,5$ and $n=5,3,2$
 respectively.  In these cases $\gamma=\frac{D-2}{2}$, and
 \be
 \beta^2=\frac{g}{D(D-2)}\alpha^{\frac{4}{D-2}},
 \ee
 where $\alpha$ is some arbitrary real-valued constant. By Wick
 rotation $t\to it$ one obtains the solutions of wave equation
 in Minkowski space-time. Using the map
 $\phi\to\Phi=t^\frac{d-1}{2}\phi$, one can also obtain the corresponding
 $SO(d)$-invariant solutions in Euclidean $\mbox{AdS}_{d+1}$ space and
 $\mbox{dS}_{d+1}$ space. For example, a solution for the
 Klein-Gordon equation for $\phi^4$ model in the ${\cal O}^-$ region of
 $\mbox{dS}_4$ space is
 \be
 \phi^-_{\mbox{dS}_4}(t,\vec x)=\frac{\alpha
 t}{\left(\frac{\alpha^2}{8}g-t^2+\left|\vec x\right|^2\right)},\hspace{1cm}\alpha\in R.
 \label{d5}
 \ee
 A method to obtain solutions given in Eq.(\ref{d3}) and more such
 solutions for conformally coupled models is as follows. Using the
 solutions $\phi_s(r)$ given in Eq.(\ref{d2}) one can try to solve the Klein-Gordon
 equation for $\phi(r)=\phi_s(r)\eta(r)$. The resulting equation for
 $\eta(s)$ is
 \be
 \left(r\frac{d}{d r}\right)\left(r\frac{d}{d r}\right)\eta+
 \left\{\begin{array}{cc}4\eta(\eta-1)=0,&
 n=2, \ d=5,\\
 \eta(\eta^2-1)=0,& n=3,\ d=3,\\
 \frac{1}{4}\eta(\eta^4-1)=0,&n=5,\ d=2.
 \end{array}\right.
 \label{d6}
 \ee
 The solutions given in Eq.(\ref{d3}) are obtained by solving
 Eq.(\ref{d6}) with vanishing constant of integration.
 \subsection{The 't Hooft ansatz and the information geometry
 metric}\label{'t Hooft}
 There exists a useful and interesting connection between the
 $SU(2)$ Yang-Mills theory and the scalar $\phi^4$ theory \cite{Actor}. This connection is a specific ansatz
 for the YM potential $W^a_\mu$ in terms of a scalar field. The
 ansatz was discovered by 't Hooft in connection with the instanton
 problem \cite{tHooft1,tHooft2}. The 't Hooft ansatz for the YM potential is,
 \be
 W^a_\mu=\eta^a_{\mu\nu}\rd^\nu\phi/\phi,
 \label{thooft1}
 \ee
 where $\eta^a_{\mu\nu}$ are the 't Hooft tensors.
 In this ansatz the equation of motion for pure $SU(2)$ gauge
 theory, reduces to the following equation
 \be
 1/\phi\rd_\mu\Box\phi=(3/\phi^2)\rd_\mu\phi\Box\phi,
 \label{thooft2}
 \ee
 which can be integrated once to give
 \be
 \Box\phi+\lambda\phi^3=0,
 \label{thooft3}
 \ee
 where $\lambda$ is an arbitrary integration constant. The $\p$ solution in four dimension corresponds to $k=1$ instanton.
 In fact, the $SU(2)$ instanton density is,
 \be
 \mbox{tr} F^2\sim\frac{\beta^4}{(\beta^2+(\vec x-\vec a)^2)^4}\sim\p^4.
 \label{thooft4}
 \ee
 In \cite{Blau} $\beta$ in Eq.(\ref{thooft4}) is considered as the size of the
 instanton, suggesting to call $\beta$ the size of $\p$.

 In D-dimensions, considering $\theta^I=\beta,a^\mu$, $I=0,\cdots,D$ in
 $\p$  as the moduli, the Hitchin information metric of the moduli space, defined as follows \cite{Hit}:
 \be
 {\cal G}_{IJ}=\frac{1}{N(D)}\int d^Dx
 {\cal{L}}_0\partial_I\left(\log{\cal{L}}_0\right)
 \partial_J\left(\log{\cal{L}}_0\right),
 \ee
 can be shown to describe Euclidean $\mbox{AdS}_{D+1}$ space:
 \be
 {\cal G}_{IJ}d\theta^Id\theta^J=\frac{1}{\beta^2}\left(d\beta^2+d
 a^2\right).
\label{Mod-met}
 \ee
 $N(D)$ is a normalization constant,
 \be
 N(D)=\frac{D^3}{D+1}\int d^D x {\cal L}_0,
 \label{N}
 \ee
 and
 \be
 {\cal{L}}_0=-\frac{1}{2}\phi\nabla^2\p-\frac{g}{\left(\frac{2D}{D-2}\right)}
 \p^{\frac{2D}{D-2}}
 =\frac{g}{D}\p^{\frac{2D}{D-2}},
 \label{L0}
 \ee
 is the Lagrangian density of the $\phi^4$ theory calculated at $\phi=\p$ \cite{Solitons}. See appendix \ref{Hitchin} for details.
 \par
 $\p$ as a function of $\theta^I$'s is a free stable-tachyon field
 on ${\mbox{EAdS}_{D+1}}$
 as it satisfies the Klein-Gordon equation given in terms of the metric (\ref{Mod-met}),
 \be
 \left(\beta^2\partial_\beta^2+(1-D)\beta\partial_\beta+\beta^2\partial_a^2+
 \frac{D^2-4}{4}\right)
 \p=0.
 \ee
 The tachyon is stable as far as $\frac{-D^2}{4}<m^2<0$
 \cite{Witten}.
 \par
 $\p$ as a function of $g$ the coupling constant (or $\beta^2$), can not be
 analytically continued to $g=0$. For $g=0$, $\p$ is the
 Green function of the Laplacian  operator i.e. $\nabla^2\phi(x,a)=\delta^D(x-a)$
 and does not satisfy the Klein~Gordon equation $\nabla^2\phi=0$ for free scalar theory.
 This shows that $\p$ can not be
 obtained by perturbation around $g=0$. In \cite{Blau}, the
 same asymptotic behavior for the instanton density is observed
 and  $\mbox{tr}F^2$ is interpreted as the {\em boundary to bulk}
 propagator of a massless scalar field on ${\mbox AdS}_5$.
 \subsection{The ansatz of conformally flat metrics}\label{Einstein}
  In this section we show that in four dimensions the
  Einstein-Hilbert action for conformally flat ansatz of metrics given by
  \be
  g_{\mu\nu}=\phi^2(x)\eta_{\mu\nu},
  \label{cf1}
  \ee
  in which $\eta_{\mu\nu}$ is the Minkowski metric, reduces to the critical scalar theory in $D=4$.
  The Levi-Civita connection in ansatz (\ref{cf1}) is given by
  \be
  \Gamma^\mu_{\nu\rho}=
  \omega_{,\rho}\delta^\mu_\nu+\omega_{,\nu}\delta^\mu_\rho-\omega_,^{\
  \mu}\eta_{\nu\rho},
  \label{cf2}
  \ee
  in which
  \be
  \begin{array}{ccccc}
  \omega=\ln\phi,&&\omega_{,\mu}=\frac{\partial}{\partial x^\mu}\omega,&&\omega_,^{\ \mu}=\eta^{\mu\nu}\omega_{,\nu}.
  \end{array}
  \label{cf3}
  \ee
  The Riemann tensor $R^{\mu}_{\nu\rho\sigma}=\left[(\Gamma^{\mu}_{\nu\sigma,\rho}+\Gamma^{\mu}_{\rho\alpha}\Gamma^{\alpha}_{\nu\sigma})-
  (\rho\leftrightarrow  \sigma)\right]$ is given by,
  \be
  R^\mu_{\nu\rho\sigma}=\left(\omega_{,\nu\rho}\delta^\mu_\sigma+\omega_{,\ \sigma}^{\
  \mu}\eta_{\nu\rho}+\omega_{,\sigma}\omega_{,\nu}\delta^\mu_\rho+\omega_,^{\ \mu}\omega_{,\rho}\eta_{\nu\sigma}+\omega_{,\alpha}\omega_,^{\ \alpha}
  \delta^\mu_\sigma\eta_{\nu\rho}\right)-(\rho\leftrightarrow\sigma).
  \label{cf4}
  \ee
  Consequently
  \be
  R_{\nu\sigma}=R^{\mu}_{\nu\mu\sigma}=(2-D)\left(\omega_{,\nu\sigma}-\omega_{,\nu}\omega_{,\sigma}+\eta_{\nu\sigma}\omega_,^{\
  \alpha}\omega_{,\alpha}\right)-\Box\omega\eta_{\nu\sigma},
  \label{cf5}
  \ee
  in  which $\Box\omega=\omega_{,\alpha}^{\ \ \alpha}$.
  Finally the scalar curvature $R=g^{\mu\nu}R_{\mu\nu}$ is given as
  follows,
  \be
  \phi^2R=2(1-D)\Box\omega+(D-2)(1-D)\omega_{,\alpha}\omega_,^{\
  \alpha}.
  \label{cf6}
  \ee
  Since $\omega=\ln\phi$, the scalar curvature can be given in terms
  of $\phi$,
  \be
  R=2(1-D)\phi^{-3}\Box\phi+(D-1)(4-D)\phi^{-4}\phi_{,\alpha}\phi_,^{\
  \alpha}.
  \label{cf7}
  \ee
  Thus in four dimensions the Einstein-Hilbert action (in units $8\pi G=1$),
 \be
 S=\frac{1}{2}\int d^Dx \sqrt{\left|g\right|}(R-2\Lambda)
 \label{cf8}
 \ee
  in ansatz   (\ref{cf1}), is equivalent to the critical scalar
  theory in four dimension (up to some boundary term) as can be seen by inserting
  $\sqrt{\left|g\right|}=\phi^D$ and $R$ from Eq.(\ref{cf7}) into
  (\ref{cf8}) for D=4,
  \bea
  S&=&\frac{1}{\xi}\int d^4x
  \left(-\frac{1}{2}\phi\Box\phi-\xi\Lambda\phi^4\right)\nn\\
   &=&\frac{1}{\xi}\int d^4x \left(\frac{1}{2}\phi_{,\alpha}\phi_,^{\
   \alpha}-\xi\Lambda\phi^4\right)-\frac{1}{2\xi}\int_\Sigma
   \phi\phi_{,\alpha}n^\alpha.
  \label{cf9}
  \eea
  where $\xi=\frac{1}{6}$ is the conformal coupling constant in four
  dimensions. The boundary term is not necessarily vanishing and its value depends on the
  asymptotic geometry of space-time. The corresponding equation of motion is,
  \be
  \Box\phi+\frac{2}{3}\Lambda\phi^3=0,
  \label{cf10}
  \ee
  which gives a solution of the Einstein equation. The validity of
  the last claim is not obvious and should be checked. The Einstein
  equation in terms of the Einstein tensor
  $G_{\mu\nu}=R_{\mu\nu}-\frac{1}{2}g_{\mu\nu}R$, is given as
  follows,
  \be
  G_{\mu\nu}+\Lambda g_{\mu\nu}=0.
  \label{cf11}
  \ee
 It is easy to verify that
 \be
 G_{\mu\nu}=4\phi^{-2}\phi_{,\mu}\phi_{,\nu}-2\phi^{-1}\phi_{,\mu\nu}-\phi^{-2}\phi_{,\alpha}\phi^{\
 \alpha}_,\eta_{\mu\nu}+2\phi^{-1}\Box\phi\eta_{\mu\nu}.
 \label{cf12}
 \ee
 The Einstein equation results in two sets of equations for $\phi$,
 \be
 \phi_{,\mu}\phi_{,\nu}=\frac{1}{2}\phi\phi_{,\mu\nu},\hspace{1cm}\mu\neq\nu,
 \label{cf13}
 \ee
 and
 \be
 4\phi^{-2}(\phi_{,\mu})^2-2\phi^{-1}\phi_{,\mu\mu}+(2\phi^{-1}\Box\phi-\phi^{-2}\phi_{,\alpha}\phi_,^{\
 \alpha}+\Lambda\phi^2)\eta_{\mu\mu}=0,
 \label{cf14}
 \ee
 where $\mu=0,\cdots,3$ and there is no summation over $\mu$.
 It is not obvious that the Fubini classical vacua $\p$  which is a (regular) solution
 of  Eq.(\ref{cf10}) (for $\Lambda>0$), also satisfies the above equations. To check it
 we first consider the ansatz $\phi=\phi(r)$ for a possible solution to Eqs.(\ref{cf13}) and (\ref{cf14})
 and assume the Euclidean space-time metric. For this ansatz Eq.(\ref{cf13}) simplifies to the following equation,
 \be
 \phi\left(\ddot\phi-\frac{\dot\phi}{r}\right)-2(\dot\phi)^2=0,
 \label{cf15}
 \ee
 where $\dot\phi(r)=\frac{\partial}{\partial r}\phi(r)$ and
 $\Box\phi=\ddot\phi+\frac{3}{r}\dot\phi$. Furthermore
 Eq.(\ref{cf14}) simplifies to a linear combination of
 Eqs.(\ref{cf15}) and the equation of motion (\ref{cf10}).
 Therefore the non-trivial check for $\phi_0$ as a solution of the
 Einstein equation is to examine the validity equation (\ref{cf15}) for
 $\phi=\phi_0$, which can be easily verified.

 Consequently the Einstein equation in four dimensions is, roughly speaking, equivalent
 to the critical scalar theory with coupling constant
 $g=\frac{2}{3}\Lambda$. This is in complete agreement with the result obtained in the next
 section in general $D$-dimensions  by a different approach see Eq.(\ref{meta12}).
 \section{Fubini vacua as the metastable de Sitter
 vacua}\label{deSitter}
 In section \ref{Fubini} it was shown that the classical vacua is
 responsible for spontaneous conformal symmetry breaking to the
 de~Sitter subgroup. Therefore it is natural to expect that the
 theory governing fluctuations around the vacua $\p$ be a
 field theory on a de~Sitter background \cite{Solitons,U1,dS}.

 Recall that in general, by inserting $\phi={\Omega}^{\frac{D-2}{4}}\pb$ and
 $\delta_{\mu\nu}=\Omega^{-1}g_{\mu\nu}$ in the action
 \be
 S[\phi]=\int d^Dx \frac{1}{2}\delta^{\mu\nu}\partial_\mu\phi\partial_\nu\phi,
 \label{meta4}
 \ee
 one obtains,
 \be
 S[\pb]=\int d^Dx \sqrt{g}\left(\frac{1}{2}g^{\mu\nu}\partial_\mu\pb\partial_\nu\pb+
 \frac{1}{2}\xi R\pb^2
 \right),
 \label{meta5}
 \ee
 i.e. a scalar theory on conformally flat background given by the metric
 $g_{\mu\nu}=\Omega\delta_{\mu\nu}$ in which  $\Omega>0$ is an arbitrary
 ${\cal C}^\infty$ function. $R$ is the scalar curvature of the
 background and $\xi=\frac{D-2}{4(D-1)}$ is the conformal coupling
 constant. For details see \cite{Ted} or appendix \ref{general aspect}.

 The conformally flat background we are looking for is the one for which the vacua $\p$ becomes a constant value. In other words
 $\Omega$ should be chosen in such a way that
 \be
 \pb_0={\Omega}^{\frac{2-D}{4}}\p=\mbox{const.},
 \label{meta6}
 \ee
 Therefore using Eqs.(\ref{fu7}) and (\ref{fu8}),
 \be
 \Omega=\pb_0\frac{D(D-2)}{g}\left(\frac{\beta}{\beta^2+r^2}\right)^2.
 \label{meta7}
 \ee
 Using Eq.(\ref{cf5}) one easily verifies that
 \bea
 R_{\mu\nu}&=&4(D-1)\left(\frac{\beta}{\beta^2+r^2}\right)^2\delta_{\mu\nu}\nn\\
 &=&\Lambda g_{\mu\nu},
 \label{meta8}
 \eea
 where the cosmological constant $\Lambda$ is defined by the following relation,
 \be
 \Lambda=\frac{g}{\xi D}\frac{1}{\pb_0}
 \label{meta9}
 \ee
 Apparently the value of the cosmological constant in this model  depends on the constant value that the classical
 vacua $\pb_0$ assumes, see Eq.(\ref{meta6}). As far as $g>0$ the cosmological constant is positive and the background is de~Sitter as
 was expected from the Fubini's result reviewed in
 section~\ref{Fubini}.
 Since,
 \be
 S[\pb]=\int d^Dx \sqrt{g}\left(\frac{1}{2}g^{\mu\nu}\partial_\mu\pb\partial_\nu\pb+
 \frac{1}{2}\xi R\pb^2-\frac{g}{\frac{2D}{D-2}}\pb^{\frac{2D}{D-2}}\right),
 \label{meta10}
 \ee
 if $\pb_0$ is required to be the minimum of the effective potential,
 \be
 V(\pb)=\frac{1}{2}\xi
 R\pb^2-\frac{g}{\frac{2D}{D-2}}\pb^{\frac{2D}{D-2}}.
 \label{meta11}
 \ee
 then from Eqs.(\ref{meta8}) and (\ref{meta9}) one verifies that,
 $\pb_0=1$ and consequently,
 \be
 \Lambda=\frac{g}{\xi D}.
 \label{meta12}
 \ee
 \subsection{Fubini vacua in 4D Minkowski space-time}\label{FRW}
 In this section we study the Fubini vacua in four dimensional
 Minkowski spacetime with metric  $\eta_{\mu\nu}=(-,+,+,+)$ and assume that $g>0$ \cite{dS}.
 \be
 \p(t,\vec x)=\sqrt{\frac{8}{g}}\frac{\beta}{\beta^2-(t-a^0)^2+\left|{\vec x}-{\vec a}\right|^2},
 \label{4d1}
 \ee
 where ${\vec x }\in {\bf R}^3$.
 Here on we assume $a^\mu=0$ for simplicity. $\p$ is singular
 on the hyperbola $t^2=x^2+\beta^2$ and we define its distance to
 an observer located on the origin to be given by $\beta$. The
 Hamiltonian density ${\cal H}$ corresponding to $\p$, is,
 \be
 {\cal H}=\frac{16\beta^2}{g} \frac{t^2+x^2-\beta^2}{(-t^2+x^2+\beta^2)^4}
 \label{4d2}
 \ee
 which tends to infinity in the vicinity of the singularity. One can show that the most stable $\p$ solution is the
 zero-sized one, corresponding to $\beta\to\infty$, see appendix \ref{stability}.
 Therefore the singularity is safe when the scalar theory is coupled to gravity.
 For $t<\beta$ one can calculate, say, the total vacuum energy $H=\int d^3x {\cal H}$ corresponding to
 $\p$ which is surprisingly vanishing, $H=0$. The de~Sitter
 background corresponding to the vacua $\p$ is given by the metric,
 \be
 ds^2=\frac{12 \beta^2}{\Lambda}\frac{1}{(\beta^2-t^2+x^2)^2}(-dt^2+d{\vec
 x}^2),
 \label{4d3}
 \ee
 where $\Lambda>0$ is the cosmological constant see Eq.(\ref{meta12}).
 This metric can be obtained using the conformal transformation $\eta_{\mu\nu}\to \Omega\eta_{\mu\nu}$
 and using Eq.(\ref{meta7})  after a Wick rotation $t\to it$.
 A different set of coordinates can be used to describe the
 corresponding  de~Sitter background  with FRW metric to see whether it is open,
 closed or flat. Defining, coordinates $u$, $\rho$ and $z_i$,
 $i=1, 2,3$ by the relations $z_i^2=1$, $t=u\cosh\rho$ and
 $x_i=u\sinh \rho z_i$ useful to describe the timelike region
 $t>\left|{\vec x}\right|$, one obtains,
 \be
 ds^2=\frac{12 \beta^2}{\Lambda}\frac{1}{(\beta^2-u^2)^2}\left(-du^2+u^2(d\rho^2+\sinh
 \rho^2 dz_i^2)\right).
 \label{4d4}
 \ee
 we define a time coordinate $\tau$ by the relation
 $d\tau=\left(\beta^2-u^2\right)^{-1} du$. Thus one obtains,
 \be
 \tau=\left\{\begin{array}{llll}
 \frac{1}{\beta}\coth^{-1}\frac{u}{\beta}&&&u>\beta,\\
 \frac{1}{\beta}\tanh^{-1}\frac{u}{\beta}&&&u<\beta,
 \end{array}\right.
 \label{4d5}
 \ee
 and
 \be
 ds^2=\frac{12\beta^2}{\Lambda}\left(-d\tau^2+\frac{\sinh^2(2\beta
 \tau)}{4\beta^2}(d\rho^2+\sinh^2\rho dz_i^2)\right)
 \label{4d6}
 \ee
 One can call the region $u<\beta$ which can be observed by observers located on the
 origin the south pole and the $u>\beta$ region the north
 pole, a known terminology in de~Sitter geometry. The south pole and north
 pole in our model are separated by the horizon located at $u=\beta$, i.e the singularity of $\p$.
 By normalizing $\tau$ by the normalization factor
 $\sqrt{\frac{12}{\Lambda}}\beta$ and defining a new coordinate
 $r=\sinh\rho$, one at the end of the day obtains,
 \be
 ds^2=-d\tau^2+a(\tau)^2\left(\frac{dr^2}{1+r^2}+r^2dz_i^2\right),
 \label{4d7}
 \ee
 in which $a(\tau)=\sqrt{\frac{3}{\Lambda}}\sinh\sqrt{\frac{\Lambda}{3}}\tau$.
 This is the Robertson-Walker metric for open de~Sitter universe.
 One can easily calculate the energy density $\rho$ and the pressure $p$ of the cosmological stuff
 corresponding to $\p$ using
 the  Friedmann equations for the open universe,
 \bea
 \left(\frac{\dot a}{a}\right)^2&=&\frac{8\pi
 G}{3}\rho+\frac{1}{a^2},\nn\\
 \frac{\ddot a}{a}&=&-\frac{4\pi
 G}{3}(\rho+3p).
 \label{4d8}
 \eea
 One verifies that $p$ and $\rho$ satisfy the equation of state
 for the cosmological constant $\rho=-p=\Lambda$ ($8\pi G=1$).
 \section{Conclusion}\label{conclusion}
 The Fubini vacua of the critical scalar theories preserves the
 de~Sitter subgroup of the full conformal symmetry. In $D$-dimensions this vacua is determined up to $D+1$ free parameters. The geometry of the
 free-parameter space is a $(D+1)$ dimensional AdS space, where the Fubini classical vacua appears as the boundary to bulk propagator.  In Euclidean spacetime,
 the entropy  of  the quantum state $\left|\Omega\right>$ corresponding to the classical Fubini vacua, given by the
 formula,
 \be
 \left<\Omega|\Omega\right>=e^{-S},
 \label{con1}
 \ee
 in 4 and 6 dimensions is given by,
 \be
 S=\left(\frac{\ell}{g^{1/4}\beta}\right)^{D-2},\hspace{1cm}D=4,6,
 \label{con2}
 \ee
 in which $g$ is the coupling constant of the scalar theory, $\ell$ is the size of the universe assumed as a spherical
 box and $\beta$ is the free parameter of the Fubini vacua.

 In  Minkowski space-time, $\beta$ appears to be equivalent to $1/T$ where $T$ is the
 temperature of the background, and one verifies that the Fubini vacua is equivalent to a bath of radiation.
 The radiation mainly consists of massless scalars with
 Rayleigh-Jeans distribution for frequencies $\omega\ll T$.

 Since the Fubini vacua is invariant under the de~Sitter subgroup of
 the full conformal group, it is reasonable to construct the
 corresponding de~Sitter background in which the Fubini vacua is a
 constant. The cosmological constant in this case is given by,
 \be
 \Lambda=\frac{g}{\xi D},
 \label{con3}
 \ee
 where $\xi$ is the conformal coupling constant. A similar result
 can be obtained in four dimensions where the Einstein field
 equation simplifies to the critical scalar theory in the ansatz of
 conformally flat metrics. In Minkowski spacetime the corresponding
 de~Sitter space is equivalent to an open FRW universe.

 A generalization of the Fubini's approach was considered in section
 \ref{dimensional reduction} by looking for a classical vacua
 with some translation symmetries. In such a vacua massless fields
 gain mass in directions along which the translation symmetry is
 broken. Thus at low-energies a lower-dimensional space with
 Poincare symmetry will be observed.
 \section*{Acknowledgement}
 The financial support of Isfahan University of Technology (IUT) is acknowledged.

 \appendix
 \section{The geometry of the moduli space in Fubini vacua}\label{Hitchin}
 Here we give a detailed calculation of Hitchin
 information metric on the moduli space of $\p$ \cite{Hit,Solitons}:
 \be
 \p=\left(\frac{D(D-2)}{g}\right)^{\frac{D-2}{4}}
 \left(\frac{\beta}{\beta^2+(x-a)^2}\right)^{\frac{D-2}{2}}.
 \ee
 From Eq.(\ref{L0}) one verifies that
 \be
 {\cal L}_0=\frac{g}{D}\left(\frac{D(D-2)}{g}\right)^{\frac{D}{2}}
 \left(\frac{\beta}{\beta^2+(x-a)^2}\right)^D.
 \ee
 Therefore
 \bea
 \partial_\beta\log{\cal
 L}_0&=&D\left(\frac{1}{\beta}-\frac{2\beta}{\beta^2+(x-a)^2}\right),\nn\\
 \partial_{a^i}\log{\cal L}_0&=&\frac{2D(x-a)_i}{\beta^2+(x-a)^2}.
 \eea
 Using these results and after some elementary calculations one
 can show that,
 \bea
 {\cal G}_{ij}&=&\frac{1}{N(D)}\int d^D x
 {\cal L}_0\partial_{a^i}\log{{\cal L}_0}\partial_{a^j}
 \log{{\cal L}_0}=\frac{4K(D)}{N(D)\beta^2}\delta_{ij}\int
 d^Dy\frac{y^2}{(1+y^2)^{D+2}},\nn\\
 {\cal G}_{\beta i}&=&{\cal G}_{i\beta}=\frac{1}{N(D)}
 \int d^D x {\cal L}_0\partial_{a^i}\log{{\cal L}_0}
 \partial_{\beta} \log{{\cal L}_0}=0,\nn\\
 {\cal G}_{\beta\beta}&=&\frac{1}{N(D)}
 \int d^D x {\cal L}_0\left(\partial_{\beta}\log{{\cal L}_0}
 \right)^2 =\frac{D K(D)}{N(D)\beta^2}\int
 d^Dy\frac{1}{(1+y^2)^D}\left(1-\frac{2}{(1+y^2)}\right)^2.
 \eea
 where $K(D)=g^{1-D/2}D^{D/2+1}(D-2)^{D/2}$.
 By performing the integrations and using Eq.(\ref{N}), one obtains,
 \be
 {\cal G}_{IJ}=\frac{1}{\beta^2}\delta_{IJ},
 \ee
 in which $\delta_{IJ}=1$ if $I=J$ and vanishes otherwise.
 \section{General properties of  conformally flat
 backgrounds}\label{general aspect}
 In this appendix we briefly review free scalar field theory in $D+1$
 dimensional (Euclidean) curved space-time \cite{Solitons,Ted}. The action for the scalar field $\phi$ is
 \be
 S=\int d^Dx\;
 \sqrt{\left|g\right|}\frac{1}{2}\left(g^{\mu\nu}\partial_\mu\phi\partial_\nu\phi+(m^2+\xi
 R)\phi^2\right),
 \label{ga1}
 \ee
 for which the equation of motion is
 \be
 \left(\Box - m^2-\xi R\right)\phi=0,\hspace{1cm}
 \Box\p=|g|^{-1/2}\partial_\mu\left(\left|g\right|^{1/2}g^{\mu\nu}\partial_\nu\p\right).
 \label{ga2}
 \ee
 (With $\hbar$ explicit, the mass $m$ should be replaced by
 $m/\hbar$.) The case with $m=0$ and $\xi=\frac{D-2}{4(D-1)}$ is referred to
 as conformal coupling.
 \par
 The curvature tensor $R^\mu_{\ \nu\rho\sigma}$ in term of Levi-Civita
 connection,
 \be
 \Gamma^\mu_{\
 \nu\rho}=\frac{1}{2}g^{\mu\alpha}(\partial_\rho g_{\alpha\nu}+\partial_\nu
 g_{\alpha\rho}-\partial_\alpha g_{\nu\rho}),
  \label{ga3}
  \ee
 is given as follows,
 \be
 R^\mu_{\ \nu\rho\sigma}=\partial_\rho\Gamma^\mu_{\ \nu\sigma}-
 \partial_\nu\Gamma^\mu_{\ \rho\sigma}+\Gamma^\mu_{\ \rho\alpha}\Gamma^\alpha_{\
 \nu\sigma}-\Gamma^\mu_{\ \alpha\sigma}\Gamma^\alpha_{\ \nu\rho}.
  \label{ga4}
  \ee
 The Ricci tensor $R_{\nu\sigma}=R^\mu_{\ \nu\mu\sigma}$ and the
 curvature scalar $R=g^{\nu\sigma}R_{\nu\sigma}$.
 \par
 The metric of a conformally flat space-time can be given as
 $g_{\mu\nu}=\Omega\delta_{\mu\nu}$, where $\Omega$ is some
 function of space-time coordinates. One can easily show that,
 \bea
 R_{\mu\nu}&=&\frac{2-D}{2}\partial_\mu\partial_\nu(\log{\Omega})-\frac{1}{2}\delta_{\mu\nu}
 \nabla^2(\log{\Omega})\nn\\&+&\frac{D-2}{4}
 \left\{\partial_\mu(\log{\Omega})\partial_\nu(\log{\Omega})-\delta_{\mu\nu}
 \delta^{\rho\sigma}\partial_\rho(\log{\Omega})\partial_\sigma(\log{\Omega})\right\},
 \label{ga5}
 \eea
 and
 \be
 \Omega R=(1-D)\nabla^2(\log{\Omega})+\frac{(1-D)(D-2)}{4}
 \delta^{\mu\nu}\partial_\mu(\log{\Omega})\partial_\nu(\log{\Omega}).
 \label{ga6}
 \ee
 By inserting $\pp=\Omega^{\frac{D-2}{4}}\pb$ in the action
 $S[\pp]=\int d^Dx \frac{1}{2}\delta^{\mu\nu}\partial_\mu\pp\partial_\nu\pp$,
 one obtains,
 \bea
 S[\pp]&=&\int d^Dx \left(\frac{1}{2}\Omega^{\frac{D-2}{2}}
 \delta^{\mu\nu}\partial_\mu\pb\partial_\nu\pb-
 \frac{1}{2}\left(\Omega^{\frac{D-2}{4}}\nabla^2
 \Omega^{\frac{D-2}{4}}\right)\pb^2\right)\nn\\
 &=&\int d^Dx \sqrt{g}\left(\frac{1}{2}g^{\mu\nu}\partial_\mu\pb\partial_\nu\pb+
 \frac{1}{2}\xi R\pb^2
 \right).
  \label{ga8}
 \eea
 To obtain the last equality the identities
 $g_{\mu\nu}=\Omega\delta_{\mu\nu}$ and $\xi\sqrt{g}R=
 -\Omega^{\frac{D-2}{4}}\nabla^2
 \Omega^{\frac{D-2}{4}}$ are used. Consequently the free massless scalar
 theory on $D$-dimensional Euclidean space, is (classically)
 equivalent to some conformally coupled scalar theory on the
 corresponding conformally flat background.
 \par
 A $D$-dimensional de Sitter (dS) space may be realized as the
 hypersurface described by the equation
 $-X_0^2+X_1^2+\cdots+X_D^2=\ell^2$. $\ell$ is called the de
 Sitter radius. By replacing $\ell^2$ with $-\ell^2$ the
 hypersurface is the $D$-dimensional anti de Sitter (AdS) space.
 (A)dS spaces are Einstein manifolds with positive (negative)
 scalar curvature. The Einstein metric
 $G_{\mu\nu}=R_{\mu\nu}-\frac{1}{2}Rg_{\mu\nu}$, satisfies
 $G_{\mu\nu}+\Lambda g_{\mu\nu}=0$, where $\Lambda=\frac{(D-2)(D-1)}{2\ell^2}$ is the
 cosmological constant.
 \section{The metastable local minima of the
 action}\label{stability}
 In this section we discuss the stability of the classical $\p$ in 4D for
 different values of the free parameter $\beta$ by studying the variation of the action by small
 field variations around it \cite{U1,dS}.

 By recasting the action in terms of new fields $\pp=\phi-\p$ one obtaines,
 \be
 S[\phi]=S[\p]+S_{\mbox{free}}[\pp]+S_{\mbox{int}}[\pp],
 \label{stab1}
 \ee
 where $S[\p]=\int d^4x {\cal L}_0=\frac{8\pi^2}{3g}$, and
 \be
 S_{\mbox{free}}[\pp]=\int d^4 x \left(
 \frac{1}{2}\delta^{\mu\nu}\partial_\mu\pp\partial_\nu\pp+\frac{1}{2}M^2(x)\pp^2\right)
 \label{stab2}
 \ee
 in which,
 \be
 M^2(x)=-3g \p^2
 =-24\frac{\beta^2}{\left(\beta^2+(x-a)^2\right)^2}.
 \label{stab3}
 \ee
 These equations show that $\p$ is a metastable local minima of the
 action. This can also be verified explicitly by numerical analysis of variations of the
 action. For this purpose it is enough to show that there are field variations
 $\p\to \phi_\eta=\p+\epsilon
 \eta$ for ${\cal C}^1$ functions $\eta$ vanishing as $x\to\infty$
 such that $\delta S=c_\eta\epsilon^2+{\cal O}(\epsilon^3)$ for some
 real positive constant $c_\eta$. For simplicity one can assume
 $\eta=\left(\frac{1}{1+x^2}\right)^n$, $g=8$, $b=1$ and calculate
 $\delta S=S[\phi_\eta]-S[\p]$ for some integers $n$. One recognizes that $c_n>0$ for
 $n>5$, though it is negative for $0<n\le 5$. A good sign for metastability
 of the action at $\p$.

 Another interesting observation is that bubbles with
 larger size are less stable than those with smaller size. This
 can be checked noting that the size of a bubble is proportional to
 $\beta^{-1}$. By repeating the above calculations one easily verifies
 that for example for $b=3$ $c_n>0$ even for $n=3$.
 Equation (\ref{stab2}) can also be used to show that the
 stability increases as $\beta\to \infty$. In fact if we calculate the variation of action
 at the stationary point $\p(\beta)$ for different values of the moduli $\beta_1$ and $\beta_2$,
 under variation $\delta\phi$, from
 Eqs.(\ref{stab2}) and (\ref{stab3}) one verifies that,
 \bea
 \Delta S&=&\delta S|_{\beta_1}-\delta S|_{\beta_2}\nn\\
 &\sim& \int d^4x
 \left(\p(\beta_2)^2-\p(\beta_1)^2\right)\delta\phi^2+{\cal
 O}(\delta\phi^3).
 \label{stab4}
 \eea
 For simplicity we assume that $a^\mu_i=0$, $i=1,2$.
 Therefore $\Delta S$ is proportional to,
 \be
 (\beta_1^2-\beta_2^2)\int_0^\infty dx
 \frac{x^3(-x^4+\beta_1^2\beta_2^2)}{(\beta_1^2+x^2)^2(\beta_1^2+x^2)^2}\delta\p^2.
 \label{stab5}
 \ee
 For $\delta\phi$ with compact support, i.e. $\delta\phi=0$ {\em
 if} $\left|x\right|>\sqrt{\beta_1\beta_2}$ the integral above is positive therefore
 $\Delta S\sim(\beta_1^2-\beta_2^2)$. As far as $\p$ is a
 metastable local minima there exist $\delta \phi$ with compact
 support such that $\delta S|_{\beta_i}>0$ $i=1,2$. Consequently if
 $\beta_1>\beta_2$ then $\delta S|_{\beta_1}>\delta
 S|_{\beta_2}>0$. One can convince herself/himself that for
 some $\delta\phi$ one obtains $\delta S|_{\beta_2}<0$
 while $\delta S|_{\beta_1}>0$. Consequently one concludes that
 there is a transition  $\beta_2\to\beta_1$ induced by say,
 thermal fluctuations. In addition the stability increases as $\beta\to\infty$.
 
\end{document}